\documentclass{WileyMSP-template}
\usepackage{subcaption}
\usepackage{adjustbox}
\usepackage{float}
\usepackage{dcolumn} % Align table columns on the decimal point
\usepackage{bm} %Bold math 
\usepackage{booktabs} % For professional table formatting
\usepackage{makecell}
\usepackage{multirow} % For multi-row cells
\usepackage[version=4]{mhchem}
\usepackage{verbatim}
\begin{document}

\justifying
\pagestyle{fancy}
\setlength{\headheight}{41.79543pt}

\title{\noindent Materials design based on a material-motif network and heterogeneous graphs}

\maketitle

% Author: Please give full first and last names for authors and include * after the name of all corresponding authors

\author{Anoj Aryal},
\author{Weiyi Gong},
\author{Huta Banjade},
\author{Qimin Yan*}

% Dedication

\dedication{}

% Affiliations: Please provide adacemic titles (Prof. or Dr.) for all authors where applicable, and include an institutional email address for all corresponding authors
\begin{affiliations}
Anoj Aryal, Weiyi Gong, Prof. Qimin Yan \\
Department of Physics, Northeastern University \\
Email Address: q.yan@northeastern.edu

Huta Banjade\\
Department of Physics, Virginia Commonwealth University\\

\end{affiliations}


\begin{abstract}

Machine learning models for functional materials design require precise and informative representations of material systems. Common representations encode atomic composition and bonding but often do not include local coordination environments across chemically diverse crystals. Recurring structural motifs provide a motif-level description of crystalline solids and can serve as interpretable descriptors for structure–property learning. To analyze the motif connectivity in materials, a bipartite material–motif network is constructed from 131,548 Materials Project entries, with materials and motifs as the two node sets. Edges connect materials to their constituent motifs and are weighted by motif distortion, which quantifies the strength of each material–motif association. Network connectivity is analyzed to identify motif-defined material clusters that capture recurring local geometries relevant to structure–property trends. Most shared motifs act as hubs that connect otherwise disconnected regions of the network, enabling motif-guided screening by expanding from known motifs to nearby materials in the same neighborhoods. A network-embedding step converts this weighted connectivity into vector representations of materials. These motif-informed embeddings enable prediction of formation energy and band gap, showing that motif connectivity provides a compact, interpretable representation that complements existing descriptors for scalable screening and structure–property modeling.

\end{abstract}

\section{Introduction}
\label{sec:introduction} 

The data-driven design and discovery of novel materials in materials science has been advanced by the availability of large-scale databases generated through high-throughput first-principles calculations \cite{jain2013commentary, curtarolo2012aflow, saal2013materials, bergerhoff1987crystallographic}. Existing materials discovery workflows adopt multi-tier screening strategies in which candidate materials are filtered based on physical descriptors such as elemental composition, crystal symmetry, and stability, followed by computationally expensive density functional theory calculations. These data-driven workflows have been used to identify candidates with promising applications in energy storage, photocatalysis, and photovoltaics. However, explicit incorporation of structural similarity between materials remains relatively uncommon, despite the strong correlation between the local atomic arrangements and material properties. Local atomic environments could therefore serve as powerful screening criteria in materials discovery, complementing property-based approaches.

Integrating such relationships into scalable workflows requires an effective framework for representing and analyzing material connectivity. Network-based approaches have shown strong potential in capturing hidden relationships in complex real-world systems, including user-location interactions in social networks \cite{lang2022poi}, transit-hub connectivity in transportation networks \cite{li2023integrating, kopsidas2024modelling}, and protein-disease networks that support targeted drug discovery in molecular biochemistry \cite{pavlopoulos2018bipartite, hostallero2022looking}. In materials science, previous studies have identified structural similarities and functional trends across materials based on features such as density of states, chemical similarities, and structure fingerprints \cite{isayev2015materials, aykol2019network, veremyev2021networks}. However, the direct use of local coordination environments in a network-driven framework remains largely unexplored. 

Local atomic environments or structure motifs describe the arrangement of atoms around a central site and form recurring building blocks in crystalline solids that strongly correlate with physical, electronic, magnetic, and optical properties. They provide a chemically interpretable and symmetry-aware basis for representing structure–property relationships. Previously, motifs have been used to explore application specific trends; for example, \ce{TiO6} octahedra in perovskites and \ce{VO4} tetrahedra in bismuth-based compounds have been associated with optoelectronic properties relevant to solar energy conversion \cite{elaouni2023bismuth, xu2023photocatalytic}. Similarly, in superconductors, Cu-O planes and \ce{CuO4} square-planar motifs in cuprates and \ce{FeAs4} tetrahedra in iron-based compounds are central to the emergence of high-$T_c$ behavior \cite{pelc2022unconventional, meng2023topotactic, stewart2011superconductivity}. Structure motif analysis has also been used to assess crystal similarity within the Materials Project database \cite{jain2013commentary, zimmermann2017assessing, zimmermann2020local, waroquiers2017statistical, ong2013pymatgen}. These examples highlight the potential of motifs as a compact yet informative representation of crystal structure.

A systematic framework for capturing the material–motif relationships can be modeled within a network graph structure where materials and motifs can be encoded as a separate set of nodes and their connectivity as edges. Network science offers a rich set of methodologies, such as random walks, centrality metrics, and graph embeddings, for analyzing both homogeneous and heterogeneous network structures. While early embedding methods focused on homogeneous networks \cite{perozzi2014deepwalk, grover2016node2vec}, many real-world systems are heterogeneous \cite{dong2017metapath2vec, huang2020biane, barabasi2004network}. In heterogeneous networks, bipartite learning approaches are effective for modeling two-type node relationships and for extracting informative low-dimensional representations for downstream tasks \cite{gao2018bine, sybrandt2019fobe, athar2023asbine}. In parallel, machine learning models, particularly graph neural networks used in crystalline solids, have shown strong predictive performance representing crystal structure as a graph with atoms as nodes and bonds as edges \cite{xie2018crystal, chen2019graph, dong2019bandgap, gong2019predicting, park2020developing, choudhary2021atomistic, ruff2024connectivity}. For better representation of motifs in crystals, motif-based crystal graph with explicitly encoded motif-level features, along with atomic and bonding information has been used as input in machine learning model for predicting electronic band gap and classifying materials as metals or non-metals \cite{banjade2021structure}. In molecular chemistry, a similar strategy has been used to construct motif-based heterogeneous graph network and generate motif-level feature representations, which enrich molecular representation \cite{yu2022molecular}.

In this work, a bipartite network connecting materials to their structure motifs is constructed, providing a systematic framework for capturing structural similarities across materials with common motifs. The connectivity encoded in the network is analyzed using the centrality metrics to identify clusters of related materials, offering a motif-centric perspective for guiding candidate selection in materials design workflows. In addition, this analysis reveals functional clusters of materials linked through characteristic motifs, such as \ce{VO4} and \ce{TiO6} in solar energy conversion, \ce{FeAs4} and \ce{CuO4} in superconductors, and \ce{NaO6}, \ce{MnO6}, and \ce{PO4} in battery electrodes. These clusters not only highlight known functional compounds but also point to other materials that show similar functionality. The results demonstrate that a motif-informed bipartite network is effective in exploring the connectivity of structurally and functionally similar materials. A bipartite network embedding method \cite{gao2018bine} is applied to learn low-dimensional vector representations of materials (and motifs). For solid state materials, the learned embeddings serve as enriched feature vectors that capture motif-based structure relationships. These representations are then used to predict key material properties such as formation energy and band gap, and metal–nonmetal classification. The learned embeddings provide an interpretable and transferable way to connect local coordination environments with material properties across chemically diverse material families. More importantly, this framework can be integrated into existing materials discovery workflows as an intermediate screening stage for identifying candidate compounds with desired structural motifs and potential functional applications.

\section{Methods}

\subsection{Structure Motif Identification}

Structure motifs are basic crystal-structure descriptors that provide chemically interpretable descriptors for structure–property relationships. For large datasets, a robust, automated procedure is required to extract motif information. Motifs are identified using a nearest-neighbor approach by analyzing each cation with its surrounding anions. The automated coordination environment analysis tool, developed by Waroquiers \textit{et al.} \cite{waroquiers2017statistical} and implemented within the ChemEnv subpackage in pymatgen \cite{ong2013pymatgen}, is used to generate the dataset for materials with motif information. However, not all identified motifs are perfect, as distortions are expected. The continuous symmetry measure (CSM) \cite{pinsky1998csm} is used to quantify the similarity of each polyhedron to the closest ideal IUPAC-defined geometry \cite{lima1990iupac}; CSM = 0 corresponds to a perfect environment, and larger values indicate increasing distortion. Motif nodes are initialized with one-hot encoding of motif type and atomic species, whereas material nodes are initialized with a one-hot encoding of atomic composition.

\subsection{Material-Motif Bipartite Network}

A bipartite graph $G(U, V, E)$, where $U$ and $V$ are the two distinct sets of nodes and $E$ is the set of edges, is defined such that $U \cap V = \emptyset$. For example, in a user-item interaction, $U$ represents the set of users and $V$ represents the set of items. Each edge $E = U \times V$ connects a node in set $U$ to a node in set $V$, indicating a particular user's interaction with a specific item. Bipartite graphs offer a unique and rich structure to represent interactions between two distinct sets of entities. In bipartite network embedding, the goal is not only to model the observed edges between the different sets of nodes but also to capture the transitive relations within the same set of nodes. 

Bipartite network embedding  is one such model to generate embedding vectors for bipartite graphs, which not only captures the relationship between two different sets of nodes defined as explicit relations but also captures transitive links within the same set of nodes, defined as implicit relations \cite{gao2018bine}. In the structure motif-based network, materials and motifs are defined as two distinct sets of nodes. An edge is created between a material and a motif if that motif occurs in the material. Each motif node is treated as an ideal motif and the geometric distortions from ideal coordination geometries are quantified using the CSM. The edge weight is defined based on CSM, assigning weight = 1 to an undistorted motif with CSM = 0 while more distorted motifs have larger CSM and therefore smaller weights. The explicit relation is defined using this edge weight, which directly links a material to its constituent motifs. For implicit relations, the neighborhood information for each node based on transitive relationships within the network is used. For each material node, the first-order neighbors of the same node type are determined by its indirect connections to other materials linked by the same motifs. These neighbors are further refined by sorting based on degree centrality to prioritize highly connected and structurally significant nodes in the graph. To reduce redundancy, previously included neighbors are excluded from subsequent contexts, allowing each material node to have a focused and non-redundant representation of its relationships. Learning to predict these relationships results in vector representations that encode both explicit and implicit patterns in the bipartite network, enabling downstream tasks such as property prediction and classification. The model jointly optimizes both relation types and generates the embeddings for both node sets in the bipartite network.

From the material-motif dataset, a bipartite network is built with one node set for materials and the other for motifs. The bipartite network is used to visualize the material space, identify patterns, and explore different material families with various applications. The networkx package \cite{hagberg2020networkx} is used to construct the network and compute centrality measures for network analyses. Gephi \cite{bastian2009gephi} and Cytoscape \cite{shannon2003cytoscape} are used for network visualization. Embedding vectors for the materials and motifs are then generated using BiNE \cite{gao2018bine}. The learned embeddings are used as input features for a neural network architecture to predict material properties (e.g., formation energy, band gap) and for classification of metals and non-metals. The predictor is a feed-forward network implemented in PyTorch \cite{paszke2019pytorch} as a two-layer multilayer perceptron with a 64-dimensional hidden layer and ReLU activation and trained using Adam optimizer. This sequential mapping enables effective feature transformation for regression and classification.

\section{Results and Discussion}
\label{sec:results}

The materials used in this study are sourced from the Materials Project (MP) database \cite{jain2013commentary}. From the processed dataset containing materials and associated motif information, a bipartite network is constructed where one node set represents materials and the other represents the structure motifs. The largest network consists of 131,548 materials from the MP database, filtered to include entries with well-defined structure motifs and no more than 50 atomic sites. In addition, two perovskite-specific networks are studied: one derived from the Matbench dataset \cite{castelli2012new} and the other from the MP database. The overall connectivity of the network can be characterized by the network density, which measures the network's total actual edges as a proportion of all possible edges. The network of materials from the MP database is sparse, with a density of 0.00003. This low density is expected because of the vast chemical and structural diversity of materials, such that each motif is compatible with only a limited subset of materials. As a result, the network exhibits only a small number of actual, physically meaningful material–motif associations out of all possible pairs.

\begin{figure}[H]
 \centering
 \includegraphics[width=0.95\textwidth]{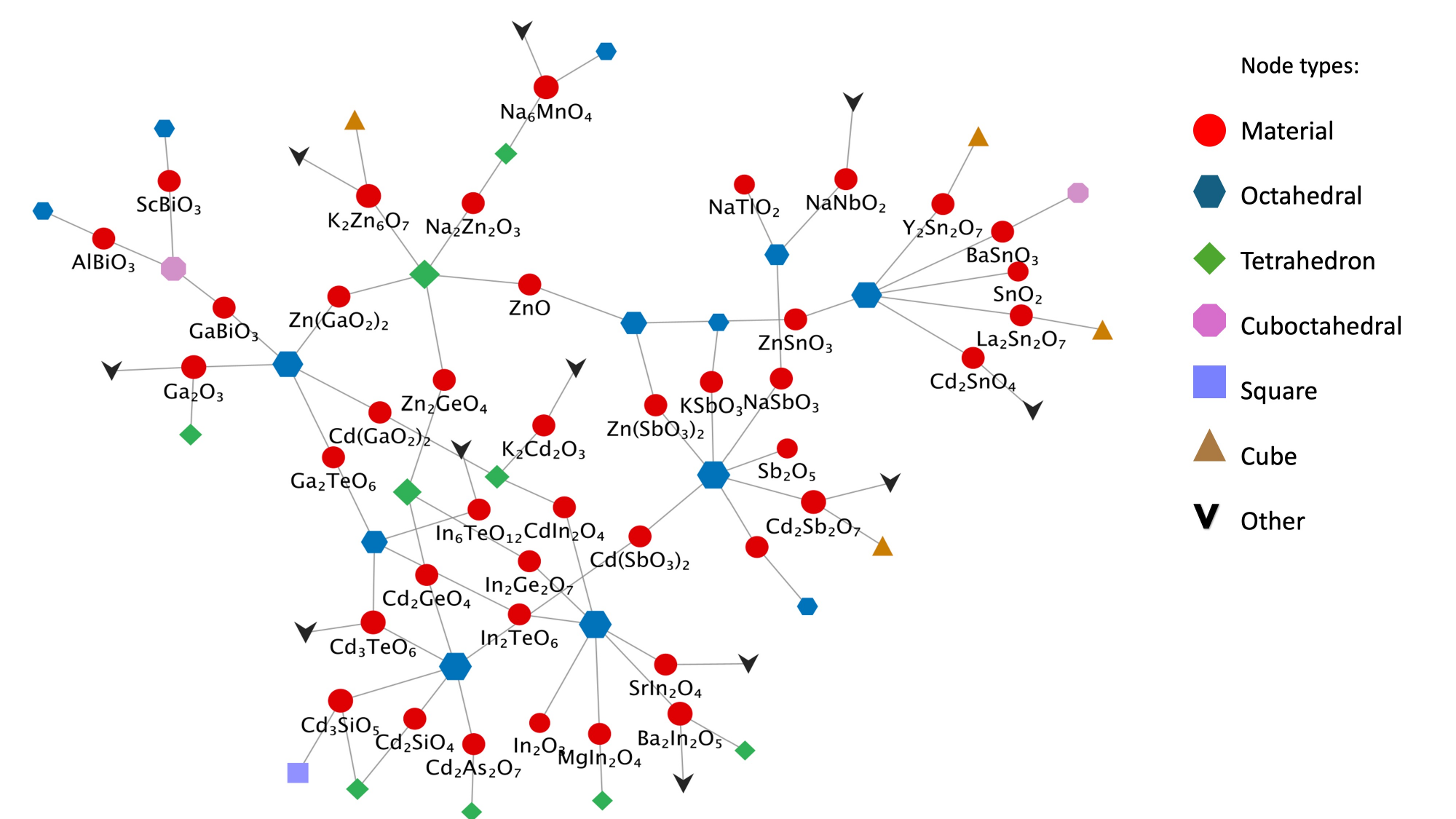}
 \caption{A sample network for a small set of complex metal oxides (red circular nodes) and their structure motifs with different motif types. Only most common motif types are shown and remaining ones are included as other types.}
 \label{fig:sample_network}
\end{figure}

In \textbf{Figure}~\ref{fig:sample_network}, a representative subgraph is presented consisting of several complex metal oxides and their motifs to visualize the structural relationships captured in the network. Shared motifs appear as hubs connecting multiple materials, revealing clusters of structurally similar compounds. Such connectivity patterns can be analyzed to identify material families with comparable local structures, which may exhibit similar functional properties. In this way, the resulting network captures explicit structural relationships by linking materials through shared motifs. To examine the structural topology of this network, various standard centrality measures are computed and analyzed.

\subsection{Centrality Measures}

Centrality measures quantify the structural importance of individual nodes within a network. In the context of a material–motif bipartite network, these metrics allow for the identification of structurally significant materials or motifs. They reveal which nodes function as hubs or bridges, thereby shaping connectivity within the network. Materials with different structures and functionalities can be linked if they share motifs. Various centrality metrics are considered to assess the network's structure and identify nodes that are central to different clusters. {Table 1} summarizes the mean centrality measures for materials and motifs in different networks studied. All centrality values shown are computed on the undirected material–motif network with projected networks of materials and motifs. The results show that the graphs are sparse, heavy-tailed, and exhibit hub–bridge structure which is similar to that observed in human–virus protein–protein interaction (HV PPI) network \cite{khorsand2020comprehensive}.

\begin{figure}[H]
 \begin{minipage}{0.33\textwidth} 
 \centering
 \includegraphics[width=\linewidth]{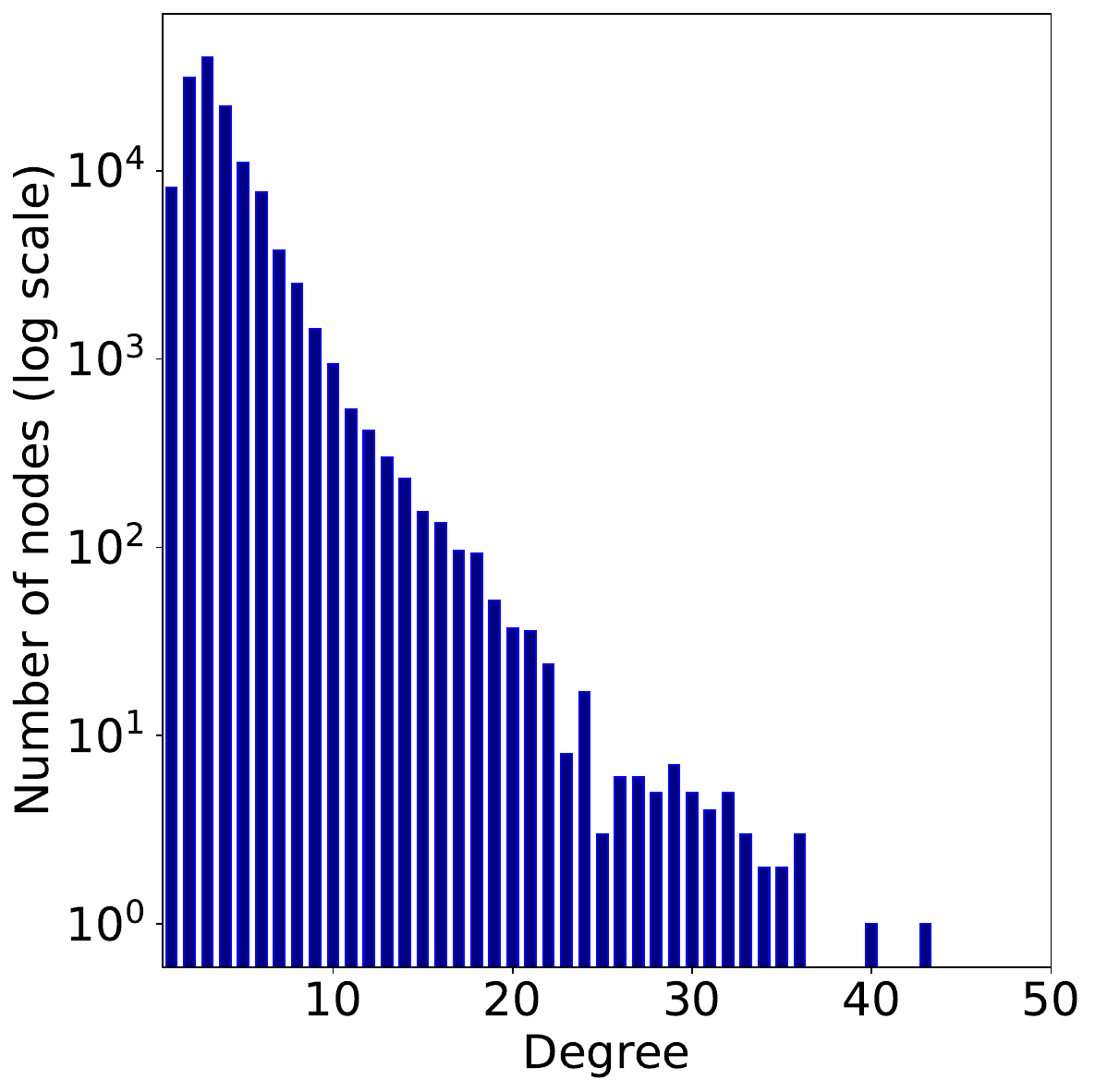}
 \subcaption{}
 \end{minipage}
 \begin{minipage}{0.32\textwidth} 
 \centering
 \includegraphics[width=\linewidth]{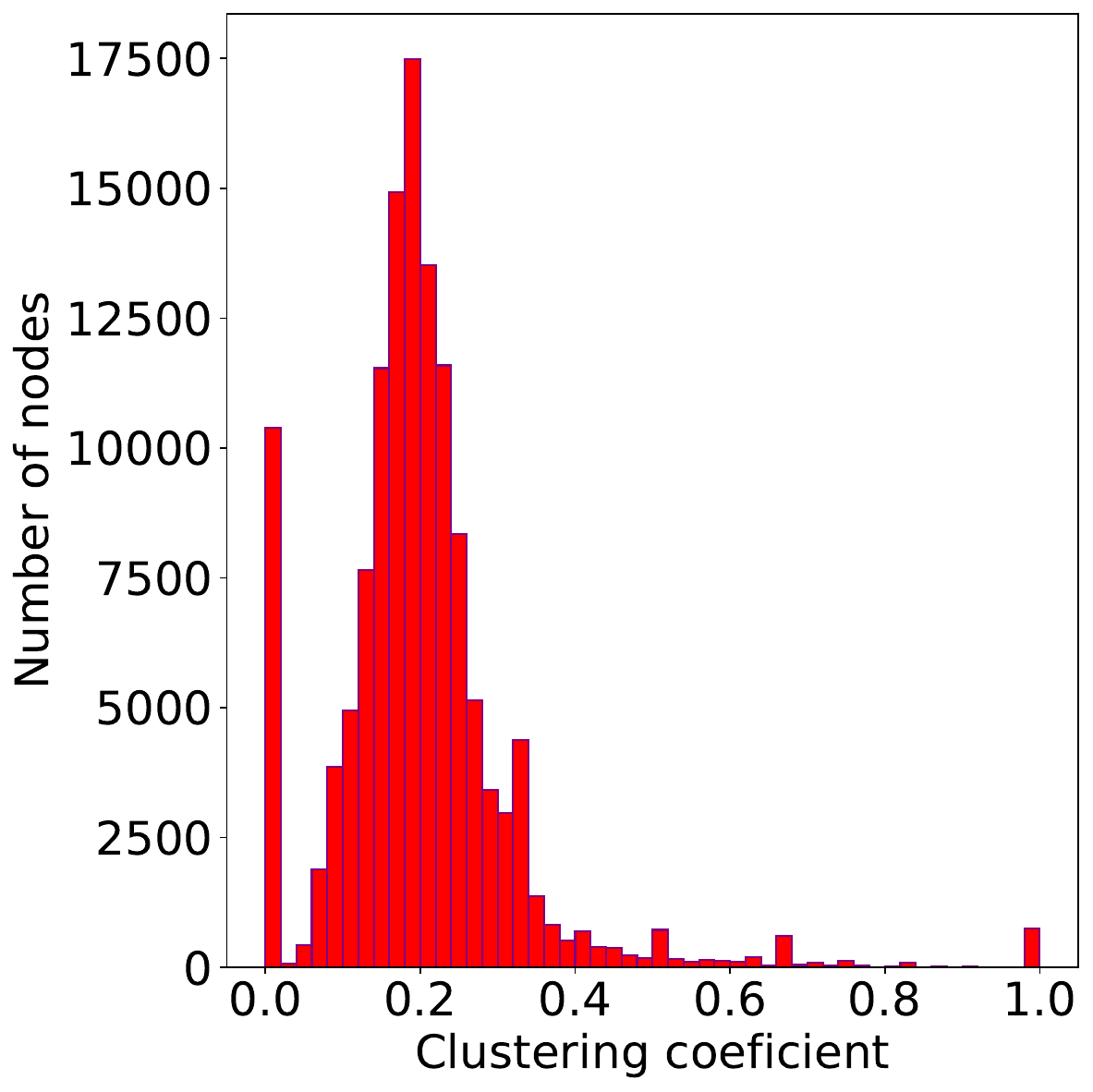}
 \subcaption{}
 \end{minipage}
 \begin{minipage}{0.32\textwidth} 
 \centering
 \includegraphics[width=\linewidth]{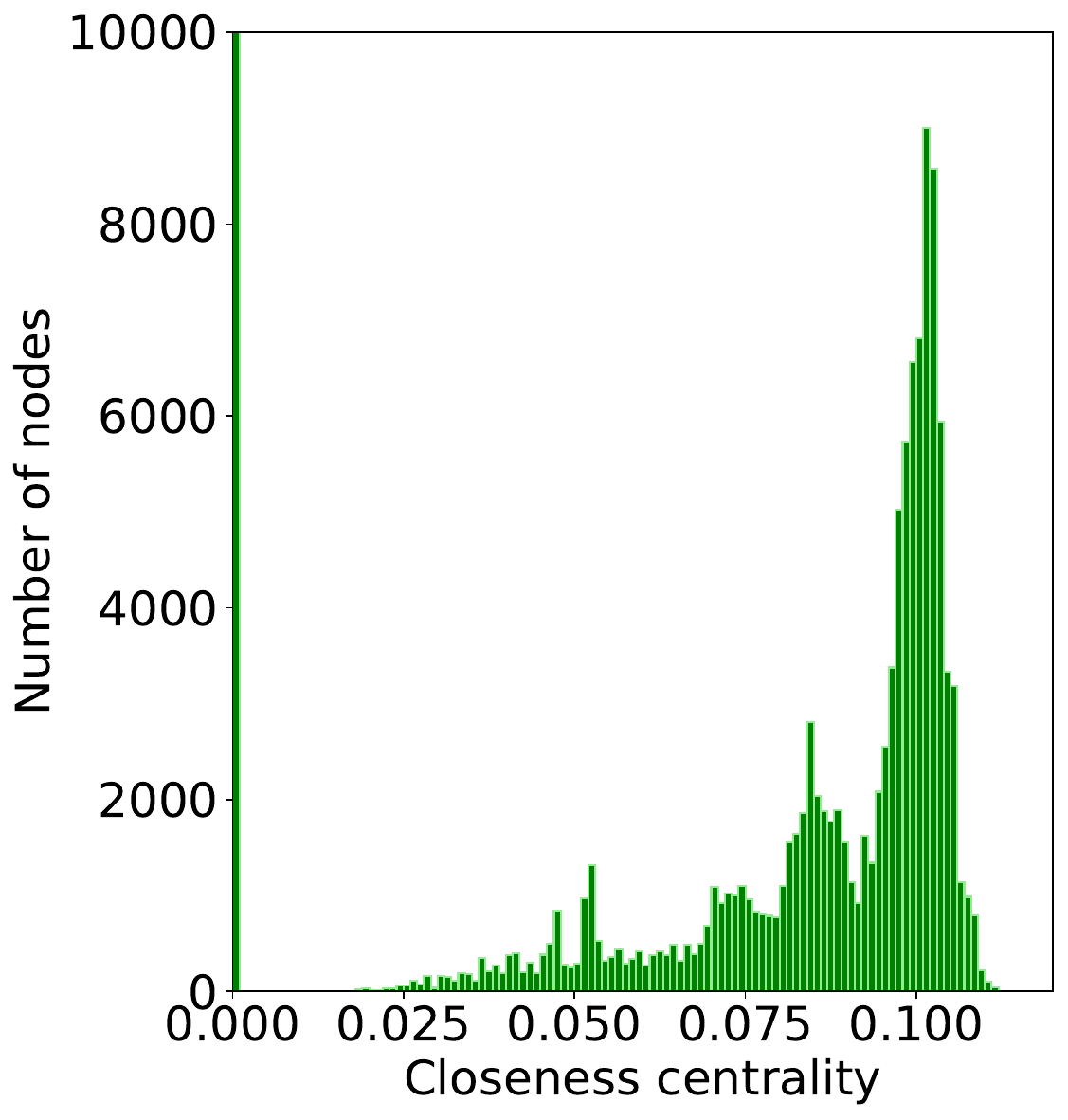}
 \subcaption{}
 \end{minipage}
 \caption{Centrality measures in network a) degree distribution, b) clustering coefficient and c) closeness centrality for material nodes in the network. Degree identifies highly recurrent motif hubs, moderate clustering reveals several local communities, and low average closeness indicates a sparse network in which only a small number of motifs and materials connect diverse material families.}
 \label{fig:centrality}
\end{figure}

\begin{figure}[H]
 \centering
 \includegraphics[width=0.95\textwidth]{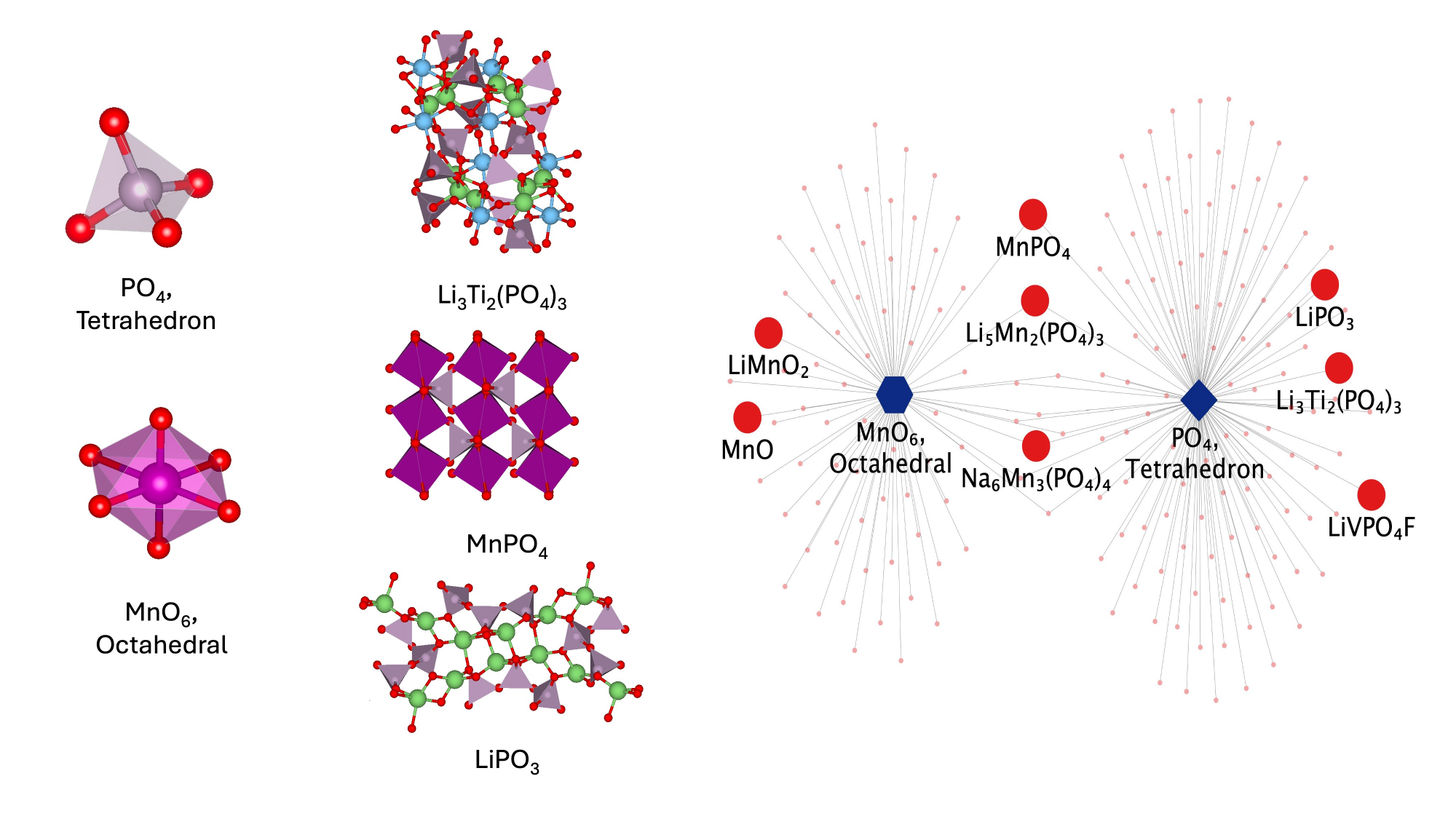}
 \caption{From left to right, the figure shows ideal local coordination motifs: \ce{PO4} tetrahedron and \ce{MnO6} octahedron, few representative crystal structures containing these motifs: \ce{Li3Ti2(PO4)3}, \ce{MnPO4}, and \ce{LiPO3}). The rightmost figure shows corresponding motif-centered subnetwork from the material–motif graph. The network panel illustrates how these motifs (blue nodes) connect multiple materials (red nodes), highlighting their role as hub motifs that link related compounds.}
 \label{fig:po4_motif_sample}
\end{figure}

Degree is a local measure of how many direct connections a node has in a graph. In a material–motif bipartite network, a material's degree is equal to the number of motifs it contains, and a motif's degree is equal to the number of materials it appears in. In large sparse networks, the distribution of degree values is heavy-tailed, approximately following a power law \cite{barabasi2013network}. As observed in \textbf{Figure}~\ref{fig:centrality}a, a small set of nodes has high degree while most nodes have small degree and connect to only a few other nodes. Degree centrality is the normalized form of degree, and identifies nodes involved in a large number of structural associations. The node with the highest degree centrality in the network is the \ce{PO4} tetrahedral motif, which is also the most common motif in the dataset. Other frequently observed octahedral motifs \ce{MnO6}, \ce{LiO6} and \ce{FeO6} also have higher degree centrality values as shown in {Table S1}. \textbf{Figure}~\ref{fig:po4_motif_sample} provides a representative example of motif-centered connectivity, showing how high-degree motifs such as \ce{PO4} tetrahedra and \ce{MnO6} octahedra connect many materials in the bipartite graph and correspond to recurring local coordination units in crystal structures. These motifs are effective in establishing connections among structurally dissimilar materials \ce{MnPO4}, \ce{MnO}, \ce{LiPO3}, and \ce{Li3Ti2(PO4)3} thereby enhancing the overall connectivity of the network.

Clustering coefficient reflects the tendency of nodes to form local communities and is defined as the ratio of the number of edges actually present among its neighbors to the number of all possible edges among those neighbors. The network has a moderately high clustering coefficient (0.201) for materials from the MP database, indicating the presence of densely connected structural nodes within local subgraphs. Nodes with high degree tend to exhibit low clustering as observed in {Table S1}, because they connect many otherwise weakly related neighbors. The clustered structure of materials in the network is visualized in \textbf{Figure}~\ref{fig:centrality}b, where the clustering coefficient histogram highlights the prevalence of locally clustered regions. For example, material \ce{MnPO4} is linked with other materials such as \ce{NaVPO5} and \ce{Li3V2(PO4)3} because of the \ce{PO4} motif and forms a material cluster with other similarly linked materials.

Similarly, nodes best positioned to influence the entire network can be identified using closeness centrality. It measures how close a node is to all other nodes in terms of shortest path lengths. Higher values indicate nodes that can efficiently interact with the rest of the network. The average closeness centrality for materials is relatively low (0.101), reflecting the sparse and fragmented nature of the network. Motifs with high closeness values are well-positioned in the network to connect different material families and help maintain the network's global connectivity. For example, materials \ce{MnWO4} and \ce{WO3} are connected through the common \ce{WO6} motif, even though they have different structures. The peak at zero closeness reflects the presence of disconnected or weakly connected nodes due to the limited motif overlap across materials.

Betweenness centrality quantifies how often a node lies on the shortest paths between other nodes, identifying potential structural bridges between otherwise disconnected clusters in the network. The low average betweenness centrality (0.0002) for materials suggests that there are few nodes acting as structural bridges between materials that are otherwise disconnected in the network. For example, material \ce{BaSnO3} acts as a bridge connecting materials sharing different motif types such as \ce{BaO12} and \ce{SnO6}. Similarly, motif \ce{MgBr6} bridges materials \ce{MgBr2} and \ce{Li2MgBr4}, which are structurally dissimilar. Such bridging materials and motifs are important in the network because they connect materials which would otherwise remain disconnected in the network.

Together, these metrics suggest that the material connectivity in the network is mediated through a small number of structurally significant motifs. Motifs such as \ce{PO4}, \ce{VO4}, \ce{MnO6}, \ce{WO6}, and \ce{BaO12} not only define material clusters, but also bridge the network connecting materials with shared motifs. This connectivity is useful for screening because local coordination environments are often associated with functional behavior; functional motifs can therefore be used to identify candidate materials for targeted applications.

\subsection{Functional Applications Revealed by Motif Connectivity}

\begin{figure}[H]
 \centering
 \setlength{\fboxsep}{2pt}
 \setlength{\fboxrule}{0.3pt}
 \newlength{\hbig}\setlength{\hbig}{0.28\textheight} % height for (a),(b)
 \newlength{\hsmall}\setlength{\hsmall}{0.20\textheight}% height for (c),(d)
 \newcommand{\insidebox}[3]{%
 \fbox{%
 \begin{minipage}{\linewidth}
 \centering
 \includegraphics[height=#1,keepaspectratio]{#2}\\[0.4em]
 {\footnotesize #3}
 \end{minipage}%
 }%
 }

 \begin{minipage}{\textwidth}
 \centering
 \insidebox{\hbig}{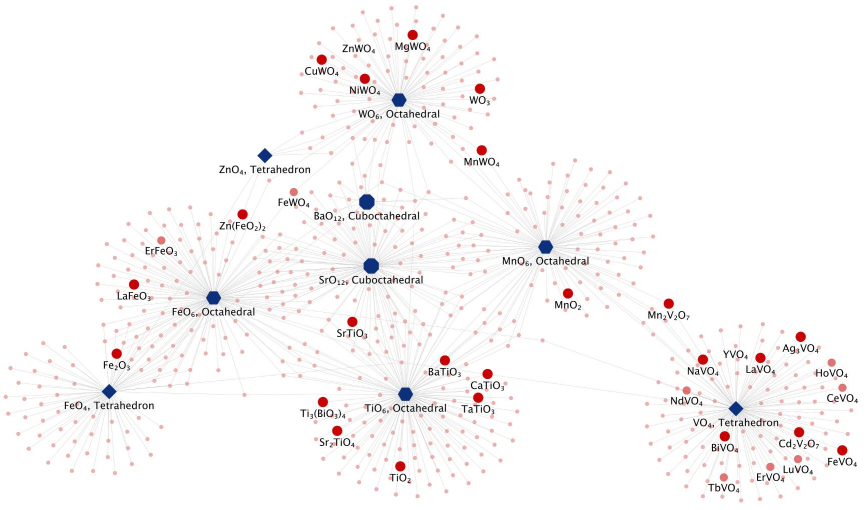}{(a) Solar energy conversion}
 \end{minipage}

 \vspace{0.6em}

 \begin{minipage}{\textwidth}
 \centering
 \insidebox{\hbig}{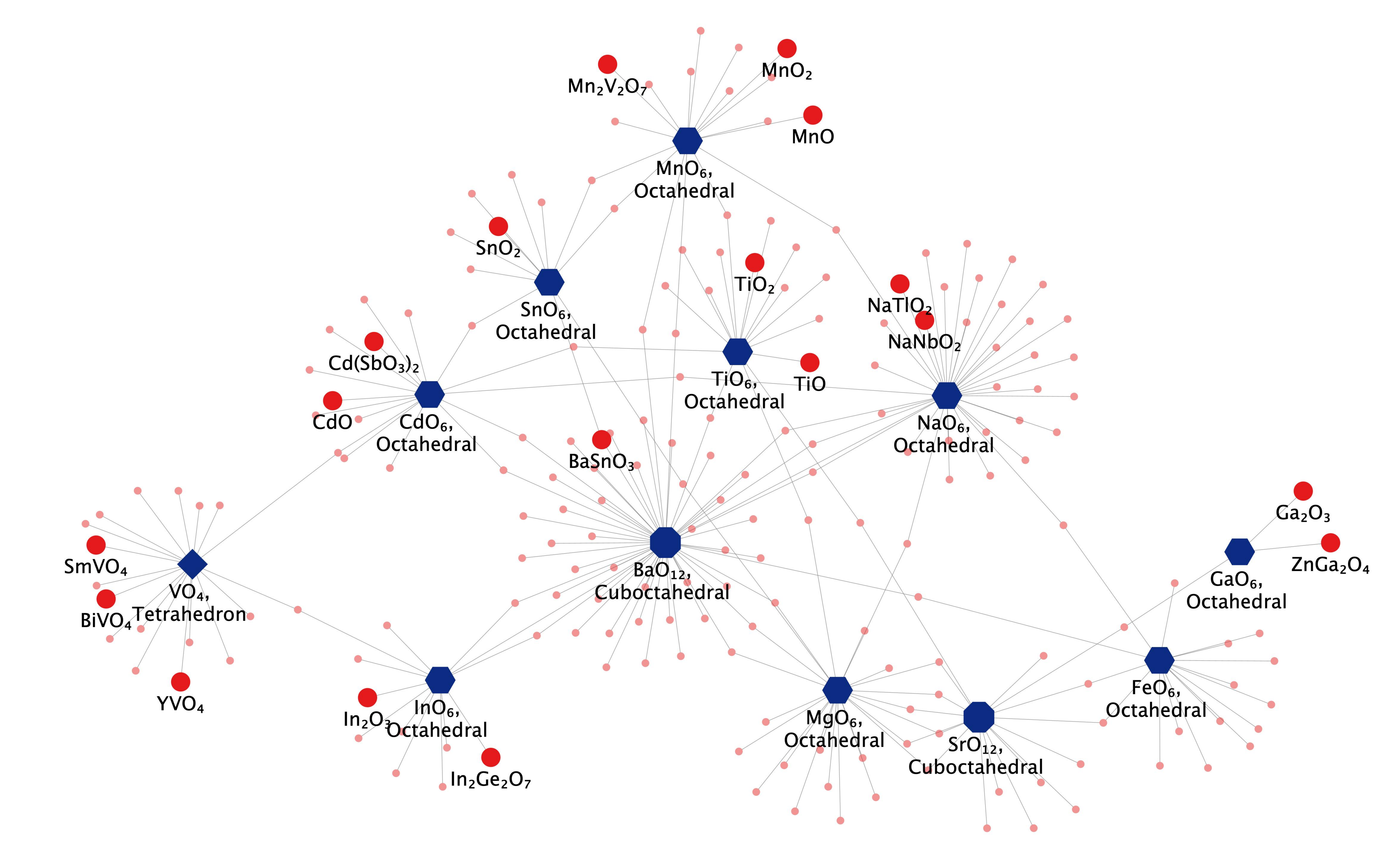}{(b) Transparent conducting oxides}
 \end{minipage}

 \vspace{0.6em}

 \begin{minipage}[t]{0.5\textwidth}
 \insidebox{\hsmall}{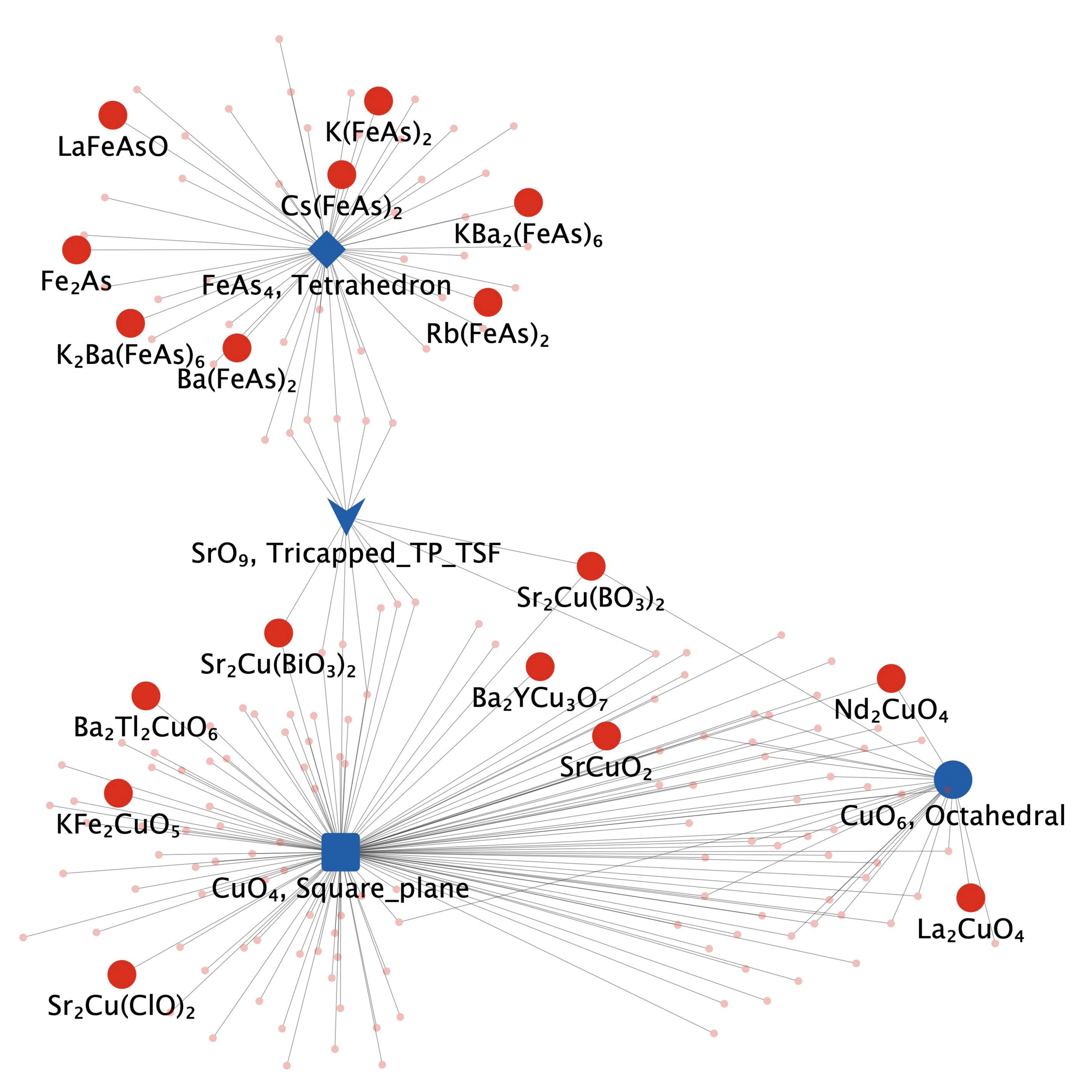}{(c) Superconductivity}
 \end{minipage}%
 \begin{minipage}[t]{0.5\textwidth}
 \insidebox{\hsmall}{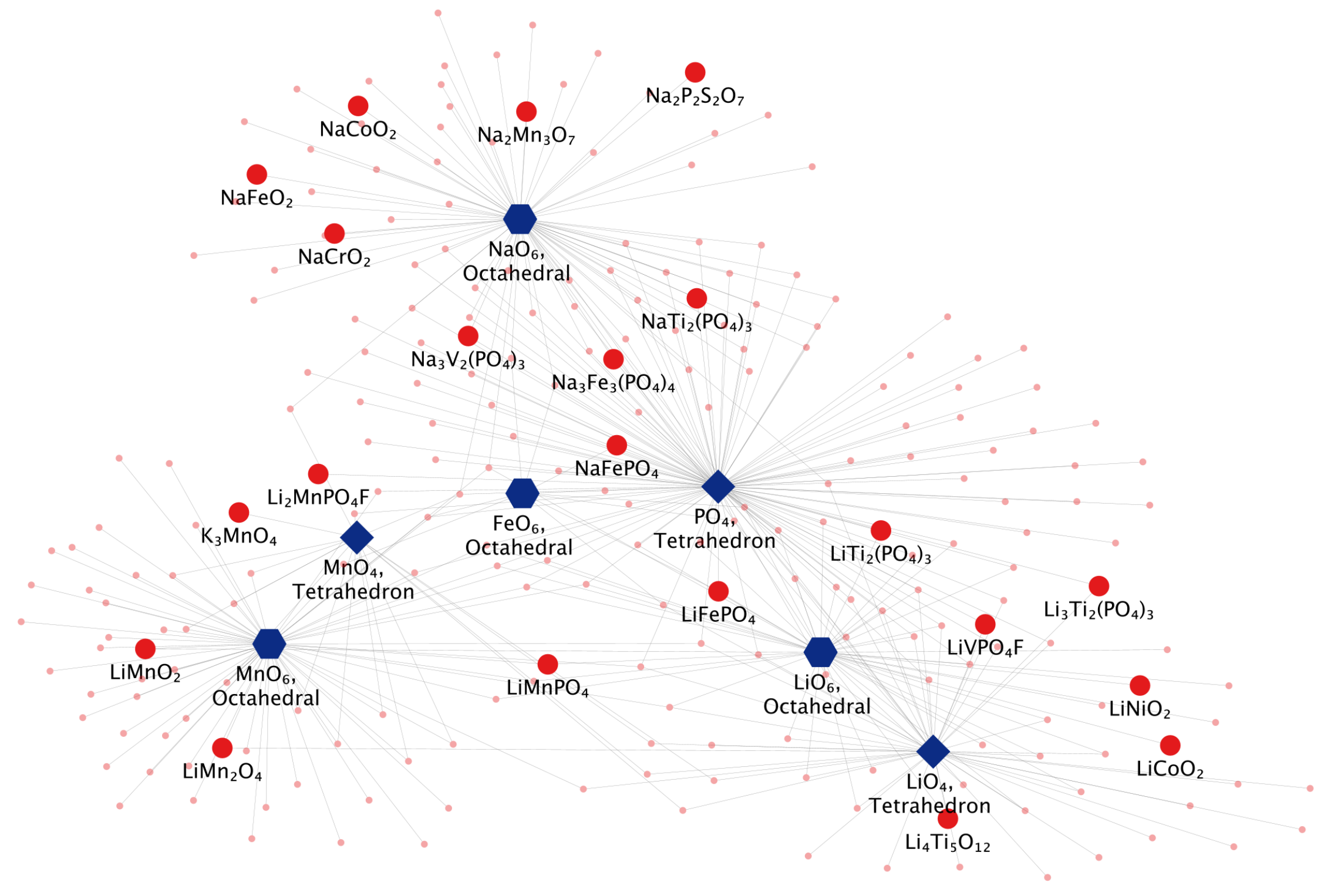}{(d) Battery electrodes}
 \end{minipage}

 \caption{Sub-networks showing material clusters based on shared motifs for
 (a) solar energy conversion, (b) transparent conducting oxides, (c) superconductivity, and (d) battery electrodes. Motif nodes are shown in blue and material nodes in red. Each subgraph highlights representative materials previously studied for the indicated functionality and their associated motifs.}
 \label{fig:application}
\end{figure}

The material–motif network provides a screening framework for identifying candidate compounds for targeted applications by exploiting connectivity through shared structural motifs. For motifs with known functionality, materials containing those motifs are collected, and an application-specific sub-network is constructed. In the material-motif network, clustering indicates that many materials share overlapping sets of motifs rather than the isolated motifs. Within each sub-network, clusters of previously studied materials are located, and their nearest neighbors within the same motif neighborhoods suggest additional candidate materials. Identified materials for solar cells, transparent conducting oxides (TCOs), superconductors, and battery electrodes are shown in \textbf{Figure}~\ref{fig:application}. 

Motifs \ce{VO4} tetrahedra, \ce{MnO6} octahedra, and \ce{TiO6} octahedra are studied for applications in solar energy conversion. Clusters of photovoltaic and photocatalytic materials emerge when these motif-based materials are linked by other functional motifs shown in \textbf{Figure}~\ref{fig:application}a (e.g., \ce{BiVO4}, \ce{$\beta$-Mn2V2O7} photoanodes and \ce{BaTiO3}, \ce{TiO2}, \ce{MnO2} photocatalysts) \cite{elaouni2023bismuth, xu2023photocatalytic, meng2023topotactic, yan2017solar, dhull2023overview, hassan2023comparative, jiao2016electronic, navarrete2025fundamentals, da2008rules, de2025titanium}. The tetrahedral \ce{VO4} motif forms a band-structure scaffold for modification by additional elements in ternary compounds. In \ce{BiVO4}, hybridization between O–\textit{2p} and Bi–\textit{6s} states perturbs the V–\textit{3d}/O–\textit{2p} manifold, tuning the valence band and band gap to values suitable for photoanode application. Occurrence of these motifs suggests that local structural units can be used as practical design rules for optoelectronic applications. The similar motif-based design rules can be used to screen candidate materials for photodetectors and transparent conducting oxides \cite{de2025titanium, o1991low, ellmer2012past, abbas2018all, saxena2024all, al2024metal, tsubota2025recent, deng20252d, nematov2024titanium, nadolska2022insight}.

TCOs are wide–band‐gap oxides that combine high electrical conductivity with visible transparency and are used in optoelectronic devices, including solar cells, light-emitting diodes, and optical sensors. In these materials, structure–property relations are governed by specific coordination motifs. For example, in the perovskite \ce{BaSnO3}, the \ce{BaO12} motif with corner-sharing \ce{SnO6} octahedra produces a highly dispersive Sn–\textit{5s} conduction band, while La substitution on Ba yields high room-temperature electron mobility \cite{jiao2016electronic, kim2012high}. Similarly, in \ce{SnO}, the Sn\(^{2+}\) \textit{5s$^2$} lone pair hybridizes with O–\textit{2p} states to generate a dispersive valence band that enables \(p\)-type transparency \cite{xu2017sn}. These structure motifs provide a clear framework for selecting materials that can be used as transparent electrodes for LEDs, photovoltaics, and sensors \cite{dhull2023overview, hassan2023comparative, jiao2016electronic, ellmer2012past, kim2012high, xu2017sn, woods2024design, morales2025synthesis, khan2020ultrasonically, zeng2007hydrothermal, wen2024advancements, brunin2019transparent, calnan2010high, kim2022deep, zhang2006growth, zhang2024recent, stadler2012transparent, walsh2009band}.

In superconducting materials, clusters of iron pnictides are centered around motif \ce{FeAs4} tetrahedral motif (e.g., \ce{LaFeAsO}, \ce{BaFe2As2}) \cite{stewart2011superconductivity, kamihara2008iron, rotter2008superconductivity, ma2008iron, sharma2021iron, dai2015antiferromagnetic}. Hybridization between transition-metal $d$ and anion $p$ states in these motifs generates correlated electronic states that are characteristic of unconventional superconductivity in superconducting materials. Similarly, in cuprates (e.g., \ce{YBa2Cu3O7}), the \ce{CuO2} planes including \ce{CuO4} square-planar units are key motifs associated with superconductivity \cite{pelc2022unconventional, hunte2008two, zhang2025additively, keimer2015quantum}. In addition, distorted \ce{CuO6} octahedra, as in \ce{La2CuO4}, strongly influence superconducting properties through octahedral tilts and bond-length distortions that modulate the Cu–O orbital overlap recognized in the discovery of high-$T_c$ cuprates \cite{keimer2015quantum, bednorz1986possible}. These motifs act as structural blocks of superconducting layers, linking together distinct material families through shared motif-driven physics. 

For rechargeable battery electrodes, motifs such as \ce{NaO6} and \ce{MnO6} octahedra, together with frameworks incorporating \ce{PO4} tetrahedra, are observed across cathode materials \cite{wang2011limn1, mcauliffe2023direct, nayak2019aluminium, gavrilova2024li3v2, wang2023design}. The octahedral environments, as found in \ce{NaFePO4}, \ce{NaCoO2}, and substituted \ce{NaMnO2} compounds, facilitate two-dimensional alkali-ion transport critical for high rate performance \cite{liu2021recent, chen2022vanadium, fu2021electrochemical}. Polyanion-based systems such as \ce{Li3V2(PO4)3} and sodium superionic conductors (NASICONs; general formula \ce{Na$_x$M2(PO4)3)}, derive their stability and higher redox potentials from the inductive effects of corner-sharing \ce{PO4} units \cite{gavrilova2024li3v2, wang2023design}. These motifs encode the essential structural principles that control ion mobility, voltage, and cycling stability across both Li- and Na-ion electrode families.

Exploring the network along such motif connections allows for screening of materials structurally analogous to known functional compounds. The shared motifs define common atomic environments that unify otherwise distinct chemical families, pointing to candidates with multifunctional or tunable properties. For materials design, motif connectivity provides a practical route to candidate discovery: starting from a motif associated with a target property, one can follow explicit \textit{material–motif–material} paths to identify additional materials that host the same structure motifs. The material–motif network is thus useful for functional materials discovery using motifs as intermediate steps in the screening process.

\subsection{Material Embeddings Based on Material-Motif Bipartite Network}

Structure motif-based bipartite networks use motif connectivity to assess material similarity. In the network representation, each material is linked to its constituent motifs, so the local coordination environments are tied directly to specific compounds. Two materials are said to be related when they share one or more motifs, though the strength of that relation depends on the geometric distortions and frequency of the motif. In the newtork, distortions are encoded as edge weights, which sharpens material–motif links beyond simple motif presence. Materials containing multiple motifs can bridge otherwise separate regions of the graph, creating transitive relations that cluster compounds with similar local geometries or functions. This motif-derived connectivity in the network can be converted into quantitative descriptors suitable for predictive modeling. The material-motif bipartite network is embedded into a vector space and the resulting material embeddings are used to assess the utility of this framework for motif-informed material descriptors and property prediction.

\begin{figure}[H]
 \centering
 \includegraphics[width=0.98\textwidth]{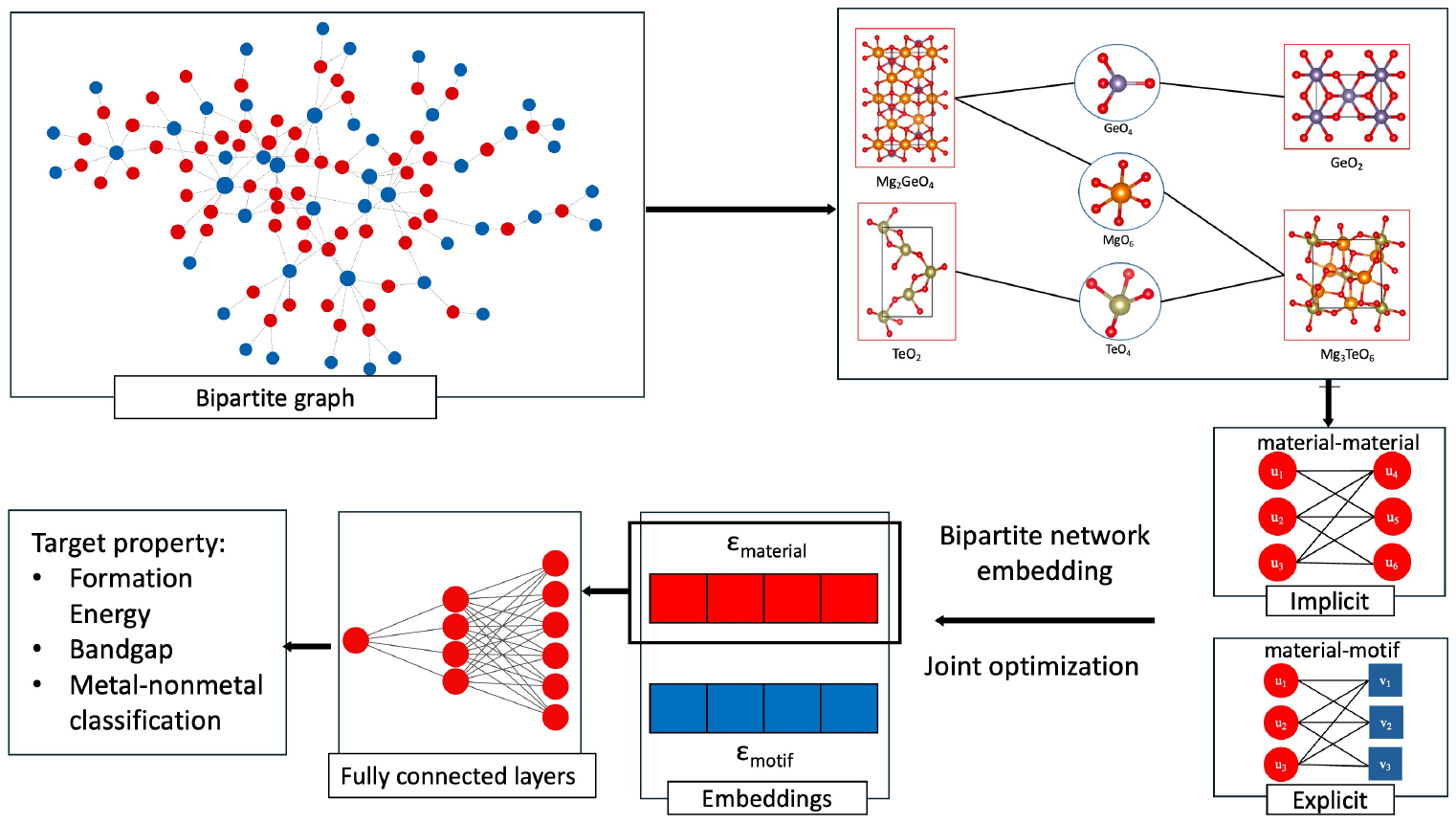}
 \caption{Schematic representation of the bipartite network embedding process and material property prediction. From material–motif bipartite graph, motif connectivity is used to explore material relationships. Embeddings are generated with BiNE using explicit and implicit objectives. The resulting material embeddings are then used for regression and classification tasks. }
 \label{fig:model}
\end{figure}

The Bipartite Network Embedding (BiNE) model \cite{gao2018bine} is adopted, which jointly optimizes two objectives: (1) preserving explicit material–motif associations via Kullback–Leibler (KL) divergence, and (2) capturing implicit transitive relations among materials mediated by shared motifs. A schematic workflow of the embedding procedure and the downstream property-prediction task is shown in \textbf{Figure}~\ref{fig:model}. Embeddings are updated using neighbor nodes under both objectives. The first-order neighbors of each material node are determined by its indirect connections to other materials linked by the same motifs. Initial node features are constructed as follows: materials are represented by one-hot atomic composition vectors, while motifs are encoded by concatenating motif type one-hot vectors with one-hot atomic vectors. These features are input to the BiNE model to learn vector representations (embeddings) of both materials and motifs. During training of explicit relations, motif site fingerprints are also used as features for motifs, as the same motif can exhibit different distortions across materials, captured by their fingerprints. The learned embeddings from the network map each node to a high-dimensional vector that captures both explicit material–motif associations and implicit transitive relationships essential for identifying structural similarity and trends in material properties. The resulting node embeddings are later used for downstream tasks such as predicting material properties. 

The learned material embeddings are used as input features for neural network models that predict formation energy, band gap, and metal–nonmetal classification. Network-generated embeddings are obtained for the materials in MP database and for perovskite materials from the Matbench dataset to evaluate predictive performance across datasets. For the regression tasks (formation energy and band gap), mean squared error loss is used for training. Train–validation–test split ratios of 70:15:15 (formation energy) and 80:10:10 (band gap) are selected for model development and evaluation, which are based on better performance metrics. In the embedding stage, the implicit objective uses first-order same-type neighborhoods (materials connected through shared motifs) to define training contexts, because these neighbors represent direct motif-mediated structural similarity. During implicit optimization, only material embeddings are updated to reflect the goal of learning material representations from motif connectivity.

After filtering the MP-derived dataset by property availability, 129,536 materials have formation energy labels and 130,548 materials have band gap labels. The Mean Absolute Error (MAE) for formation energy prediction is $0.157\,\mathrm{eV}\,\mathrm{atom^{-1}}$, and for band gap prediction $0.601\,\mathrm{eV}$ ({Table 2}), and the combined (implicit + explicit) embeddings improve over the implicit-only and explicit-only baselines. This trend indicates that both material–motif associations and motif-mediated transitive material–material relations contribute to the learned representations used for prediction. For a network generated using the Matbench perovskites dataset, which is dominated by octahedral and cuboctahedral motifs, the MAE for formation energy prediction is $0.164\,\mathrm{eV}\,\mathrm{atom^{-1}}$ ({Table 3}).

A binary classification task that classifies materials as metals or non-metals is also evaluated using the cross-entropy loss. This classification is useful for materials screening in applications that require either conducting or insulating behavior. Using the combined (implicit + explicit) embeddings, the classifier achieves 83\% test accuracy with precision(metal)=0.806, recall(metal)=0.796, and F1(metal)=0.801 (metal treated as the positive class), together with ROC–AUC=0.901. In comparison, the implicit-only and explicit-only embeddings yield lower accuracies (72\% and 74\% accuracy respectively), which highlights the importance of jointly modeling implicit and explicit relations. 

Comparing the performance metrics, embeddings from the combined relations show improved results over both implicit-only and explicit-only embeddings ({Table 2}). For the bipartite network, both the formation energy and band gap MAEs are lower for the combined relations than for the implicit and explicit baselines. This indicates that explicit material–motif links and implicit transitive relations capture complementary information: the explicit relation preserves motif-specific coordination in materials, while implicit or transitive connectivity links materials through shared motifs. In addition to representations that focus primarily on composition and bonding, the motif-informed embeddings add an interpretable description of local coordination environments that is transferable across chemically distinct material families. Although state-of-the-art crystal graph models can achieve lower errors on these benchmarks, they are primarily designed to maximize predictive performance through convolution-based learning on full crystal graphs. In contrast, the present work establishes motif connectivity as a complementary framework for material screening and structure–property learning, particularly when structural motifs are expected to play an important role in determining material properties. This framework is not only intended for property prediction, but also for interpretable materials discovery, where motif-level relationships provide a natural basis for screening and identifying candidate materials across chemically diverse families.

\section{Conclusion}
In summary, a bipartite network representation of materials and their structure motifs is introduced that captures both direct relationships between materials and their structure motifs and transitive relationships among materials through shared motifs. Centrality analysis reveals that, despite the sparsity of the network, motif-sharing patterns contribute to formation of material clusters, supporting the role of motifs as fundamental building blocks in materials. Clusters defined by functional motifs associated with different applications allow for expansion of the searchable material space to a large set of candidate materials. This framework can be used as an intermediate step in high-throughput materials discovery, enabling deeper exploration of the structure-property relationships in materials. Knowledge of functional motif driven material discovery can be a key input for designing materials with targeted properties.

Embeddings learned from the combined motif-connectivity network outperform those based only on implicit or explicit relationships in predicting formation energy, band gap, and metal–nonmetal classification. These results show that direct material–motif relationships and transitive relations among materials through shared motifs capture complementary structural information and together yield a motif-informed vector representation for structure–property learning. This representation provides an interpretable and transferable framework for relating local coordination environments to material properties across chemically diverse material families. This makes motif connectivity a useful complementary strategy for network-guided screening and the discovery of materials with targeted properties.

% Acknowledgements
\medskip
\textbf{Acknowledgements} \par %delete if not applicable))
We acknowledge funding support from the U.S. Department of Energy, Office of Science, Basic Energy Sciences, under Award No. DE-SC0023664. This research used resources of the National Energy Research Scientific Computing Center (NERSC), a U.S. Department of Energy Office of Science User Facility located at Lawrence Berkeley National Laboratory, operated under Contract No. DE-AC02-05CH11231 using NERSC award BES-ERCAP0029544.

% References
\providecommand{\noopsort}[1]{}\providecommand{\singleletter}[1]{#1}%

\newpage
% Figures/tables and captions
\section{Figures and Tables:}

\newpage
\begin{table}[H]
\centering

\caption{Centrality measures for the networks studied. The material–motif networks show moderate clustering and low average betweenness, consistent with sparse connectivity with a small set of bridging nodes.}
\vspace{0.25em}
\renewcommand{\arraystretch}{1.3}
\setlength{\tabcolsep}{14pt}
\begin{tabular}{|c c c c|}
\hline
\textbf{Network} 
& \makecell{\textbf{Clustering} \\ \textbf{coefficient}} 
& \makecell{\textbf{Betweenness} \\ \textbf{centrality}} 
& \makecell{\textbf{Closeness} \\ \textbf{centrality}} \\
\hline
Materials (MP database \cite{jain2013commentary}) & 0.201 & 0.0002 & 0.101 \\
Perovskites (Matbench \cite{castelli2012new}) & 0.300 & 0.0003 & 0.195 \\
Human–virus PPI \cite{khorsand2020comprehensive}& 0.095 & 0.0002 & 0.320 \\
\hline
\end{tabular}
\label{tab:centrality}
\end{table}

\begin{table}[H]
\centering

\caption{Property-prediction performance on Materials Project materials using learned material embeddings from material–motif network. Results are reported for combined (implicit and explicit), implicit-only, and explicit-only embeddings.}
\vspace{0.25em}
\renewcommand{\arraystretch}{1.3}
\setlength{\tabcolsep}{6pt}
\begin{tabular}{|c c c c|}
\hline
\textbf{Metric} & \makecell{\textbf{Combined} \\ \textbf{(Implicit + Explicit)}} & \textbf{Implicit} & \textbf{Explicit} \\
\hline
Formation energy MAE ($\mathrm{eV}\,\mathrm{atom^{-1}}$) & 0.157 & 0.284 & 0.341 \\
band gap MAE ($\mathrm{eV}$) & 0.601 & 0.649 & 0.673 \\
\makecell{Metal–nonmetal \\ classification accuracy (\%)} & 83\% & 72\% & 74\% \\
\hline
\end{tabular}
\label{tab:metrics}
\end{table}

\begin{table}[H]
\centering
\caption{formation energy prediction error (MAE) for perovskite subsets from the Materials Project and Matbench datasets using learned material embeddings.}
%\vspace{0.25em}
\renewcommand{\arraystretch}{1.3}
\setlength{\tabcolsep}{6pt}
\begin{tabular}{|c c c|}
\hline
\textbf{Metric} & \makecell{\textbf{Perovskites} \\ \textbf{(MP database)}} & \makecell{\textbf{Perovskites} \\ \textbf{(Matbench)}} \\
\hline

Formation energy MAE ($\mathrm{eV}\,\mathrm{atom^{-1}}$) & 0.171 & 0.164 \\
\hline
\end{tabular}
\label{tab:perov}
\end{table}

\begin{figure}
\textbf{Summary}\\
\medskip
 \includegraphics[width=1.0\textwidth]{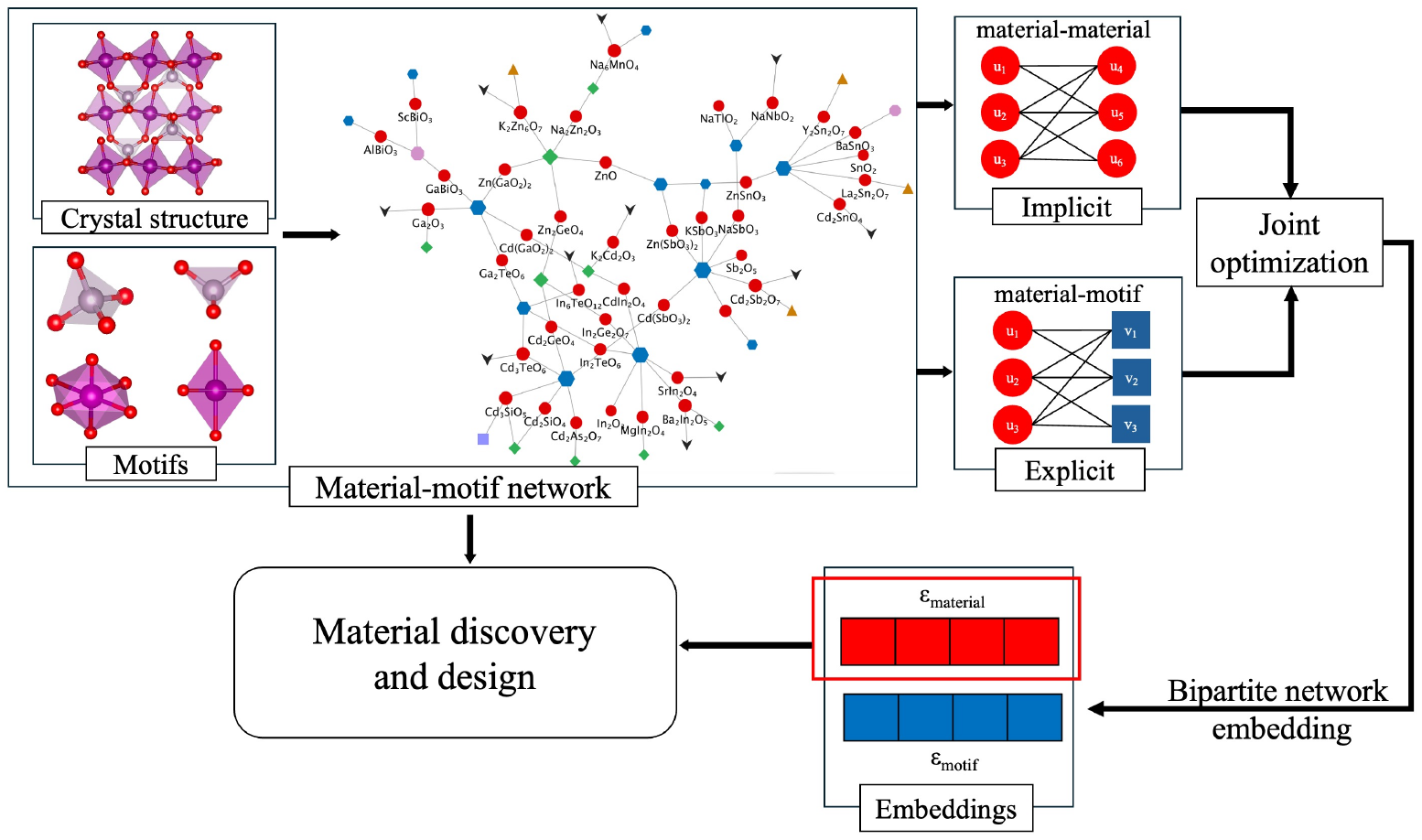}
 \medskip
 \caption*{Structure motifs in materials are used to construct a bipartite material–motif network that links each material to its constituent motifs and establishes connectivity among materials sharing common motifs. Network analysis reveals material clusters associated with different functional applications and supports motif-guided screening of materials. The bipartite graph can also be used directly for motif-guided identification of candidate functional materials, while network embedding converts motif connectivity into feature representations, yielding motif-informed descriptors for materials that enable property prediction and screening.}
 
\end{figure}

\end{document}